\documentclass[%
 aip,
 amsmath,amssymb,
 reprint,
]{revtex4-2}

\usepackage{graphicx}
\usepackage{dcolumn}
\usepackage{bm}

\usepackage[utf8]{inputenc}
\usepackage[T1]{fontenc}
\usepackage{mathptmx}
\usepackage{etoolbox}
\usepackage{makecell}

\usepackage{amssymb}
\usepackage{amsmath}

\usepackage[normalem]{ulem}
\usepackage{times}
\usepackage{graphics}
\usepackage{bm}
\usepackage{subfigure}
\usepackage{graphicx}
\usepackage{amsthm}
\usepackage{notes2bib}
\usepackage{amssymb}
\usepackage{url}
\usepackage{enumerate}
\usepackage{epstopdf}
\usepackage{mathtools}          
\usepackage{setspace}
\usepackage{color}

\makeatletter
\def\@email#1#2{%
 \endgroup
 \patchcmd{\titleblock@produce}
  {\frontmatter@RRAPformat}
  {\frontmatter@RRAPformat{\produce@RRAP{*#1\href{mailto:#2}{#2}}}\frontmatter@RRAPformat}
  {}{}
}%
\makeatother
\begin{document}

\preprint{AIP/123-QED}

\title[Geometric structure of ideal data-driven dynamical model
using RfR method]{
Geometric structure of ideal data-driven dynamical model
using RfR method
}
\author{Natsuki Tsutsumi}
\affiliation{Graduate School of Business Administration, Hitotsubashi University, Tokyo 186-8601, Japan}
\author{Kengo Nakai}
\affiliation{Graduate School of Environment, Life, Natural Science and Technology, Okayama University, Okayama 700-8530, Japan\\
(Electronic mail: t.natsu2653@gmail.com)}
\author{Yoshitaka Saiki} 
\affiliation{Graduate School of Business Administration, Hitotsubashi University, Tokyo 186-8601, Japan}

\date{\today}

\begin{abstract}
The Gaussian radial function-based Regression (RfR) method is a data-driven modeling approach that utilizes physically understandable variables from scalar time series, constructed using delay coordinates and Gaussian radial basis functions.
Even when a model successfully describes an approximate trajectory of the original system, data-driven models rarely reconstruct negative Lyapunov exponents of chaotic dynamics.
An ``ideal model'' should reconstruct the dynamical structure, including the negative (physically dominant) Lyapunov exponents.
Comparing the ideal model and the non-ideal model, we investigate the geometric structure of the attractor of such models using the Lyapunov exponents and the corresponding Lyapunov vectors.
Our investigation suggests that the ideal model reconstructs the original system's attractor as a time-delay embedding.
By applying the results, we search for a method to construct an ideal model, which persists against the change in hyperparameters.
\end{abstract}

\maketitle

\begin{quotation}
Data-driven modeling methods for time-series data, utilizing delay coordinates and Gaussian radial basis functions, have garnered attention in various fields over the past several decades.
In data-driven dynamical modeling, the goal is not only to predict short-term trajectories but also to reproduce the underlying dynamics and structure of the system. 
This paper focuses on the geometric properties of the constructed models, demonstrating how they reconstruct the original dynamics. 
Based on these properties, we propose a method to design models that capture dynamics with high precision.
The results enhance the quality of modeling and broaden its potential applications across diverse scientific and engineering domains.
\end{quotation}

\section{Introduction}
Researchers are eager to identify a dynamical system model that generates observable time series data showing chaotic behavior. 
Various methods have been proposed to estimate a dynamical system describing observed data, combining physical knowledge and machine learning techniques~\cite{chorin2015,karniadakis2021,raissi2019,greydanus2019}.
Several approaches concern modeling dynamics using machine learning from given time series data without physical knowledge.
One method, based on the theory of the Koopman operator~\cite{lin2021, mezic2013, berry2015, proctor18}, projects the phase space into an infinite-dimensional space for constructing a model as a linear system.
Furthermore, the dynamic mode decomposition~\cite{schmid2010,proctor16} is a method for deriving macroscopic features and estimating the dynamics of a dynamical system from its variables.
Reservoir computing~\cite{Pathak_2017,kobayashi2021} is a recurrent neural network with a low computational cost.
Neural ordinary differential equations~(ODEs)~\cite{chen2018} recently attracted much attention for using a neural network to model ODEs. 
In many cases, these methods generate a high-dimensional, physically incomprehensible space.

To construct a model using only physically understandable variables and space, the delay coordinates~\cite{takens_1981} and the Gaussian radial basis function~\cite{bishop2006,kawano2007} are useful.
Over the past several decades, a method for constructing a map that describes discrete time series using them has been studied and applied in various fields~\cite{smith1992,small2002}.
For continuous-time dynamics, a method for constructing an ODE model using delay coordinates and the Gaussian radial function~\cite{tsutsumi22} has been recently proposed, based on simple regression methods~\cite{wang2011, brunton16}.
This family of methods is called the radial function-based regression (RfR) method.
Some studies~\cite{tsutsumi22, tsutsumi23,tsutsumi24} report that the RfR method works well regarding not only the short-term predictability of a trajectory but also the reconstruction of long-term trajectory characteristics, such as the density distribution.
The RfR method functions even when observable variables are obtained from an infinite-dimensional system, such as partial differential equations and delay differential equations~\cite{tsutsumi23}.
When the targeted variable's behavior is complex, including the high-frequency intermittent behavior of a fluid flow, we can apply the joint RfR method~\cite{tsutsumi24} using the targeted variable and another variable (base variable) that exhibits relatively simple behavior.
The method constructs two models.
The first is an autonomous system of the base variable, and the second concerns the targeted variable being affected by a term involving the base variable, demonstrating complex dynamics.

A data-driven model is often evaluated by how accurately and how long the model can predict a short-term trajectory.
For a data-driven model to mimic the original dynamics,
the model should reconstruct the invariant sets of the original dynamics and their characteristics, such as the attractor, the invariant density distribution, and the Lyapunov exponents.
Previous studies~\cite{tsutsumi22, tsutsumi23, tsutsumi24} report that a data-driven model constructed using the RfR method can reproduce the attractor, the density distribution, and the non-negative Lyapunov exponents.
Even if the constructed models generated a trajectory that mimics an actual trajectory, they do not always reconstruct the negative exponents.

We focus on the first $N$ Lyapunov exponents to evaluate a constructed model for a $\mathcal{D}$-dimensional dynamical system in terms of the Lyapunov exponents.
Here, $N\in{\mathbb N}$ is the smallest value such that $\sum^N_{i=1}\Lambda_i<0$ and $\Lambda_i$ is the $i$-th Lyapunov exponent~($i=1,\ldots, \mathcal{D}$) of the original dynamics. 
The Lyapunov exponents $\{\Lambda_i\} (i=1,\ldots,N)$ of the $\mathcal{D}$-dimensional chaotic dynamical system ($N\le \mathcal{D}$) are called the {\bf physically dominant Lyapunov exponents}.
The physically dominant Lyapunov exponents primarily determine the dynamics.
The Kaplan-Yorke formula uses the $N$ Lyapunov exponents to estimate the Hausdorff dimension of the chaotic attractor~\cite{kaplan_1979}.
Some studies directly estimate the physically dominant Lyapunov exponents from an observable time series without constructing a dynamical system~\cite{wolf1985,parlitz_1992,brown_1991,sun2012}.
A data-driven model ideally reconstructs the physically dominant Lyapunov exponents, including some negative exponents.
We construct a data-driven model that reconstructs all of the physically dominant Lyapunov exponents of the original discrete or continuous-time dynamics using the RfR method.
The details of the data-driven modeling method are presented in the following section.

\begin{figure*}[ht!]
    \begin{center}
        \includegraphics[width=0.8\linewidth]{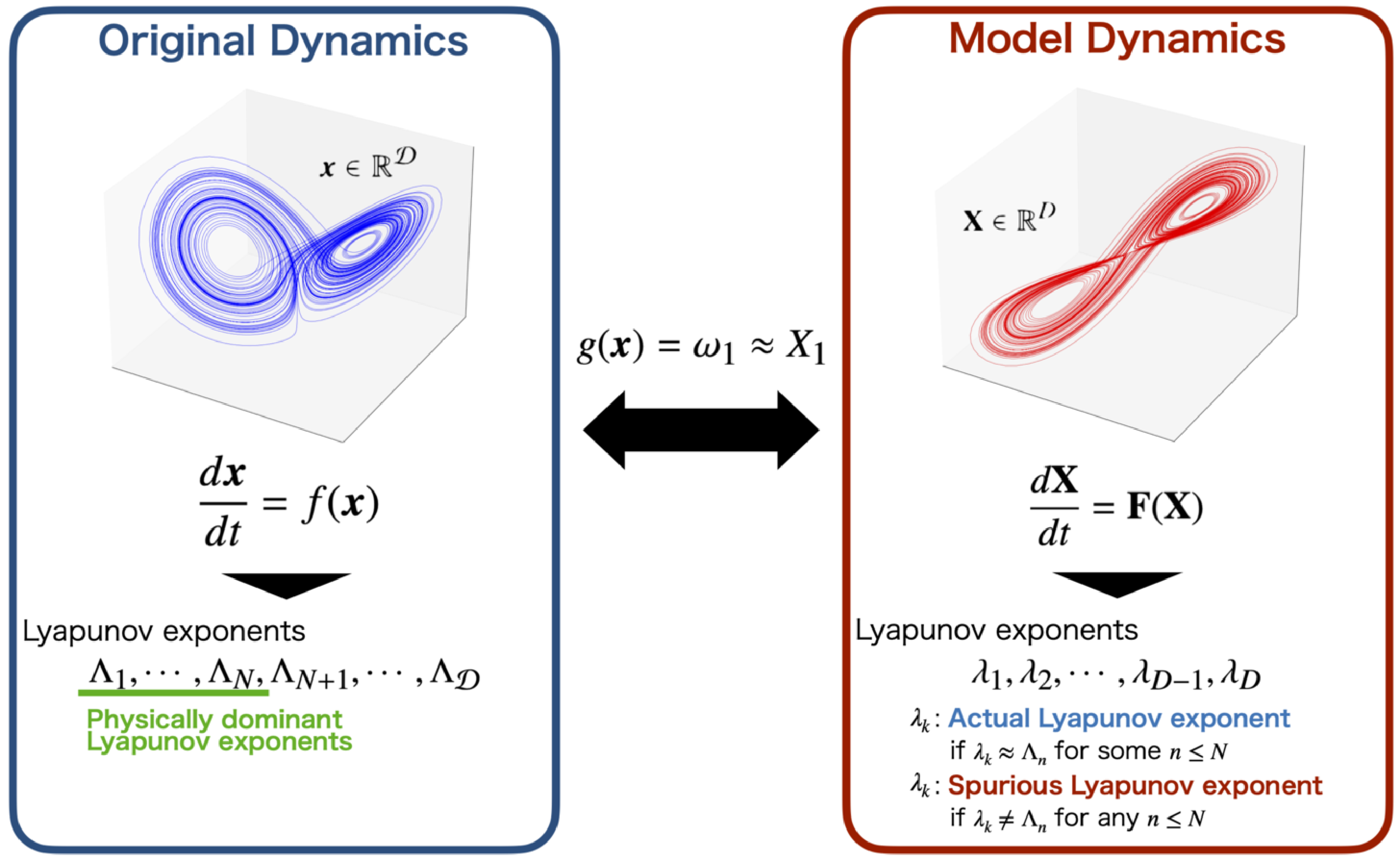}
    \end{center}
    \caption{
    {\bf Schematic picture of the Lyapunov exponents' definitions for the original and model dynamics.} 
    For the original dynamics, the first $N(\leq \mathcal{D})$ Lyapunov exponents are called the physically dominant Lyapunov exponents, which should be reconstructed in a data-driven model.
    For an ideal data-driven model, the actual Lyapunov exponents are the model's Lyapunov exponents, which are among the physically dominant Lyapunov exponents of the original dynamics.
    The other Lyapunov exponents of the model are referred to as the spurious Lyapunov exponents.
    } 
    \label{fig:lyap-img}
\end{figure*}

We define an {\bf ideal model} by a data-driven model constructed using the RfR method that satisfies the following three conditions:
\begin{itemize}
    \item [(c0)] The model reconstructs the non-negative (positive and neutral) Lyapunov exponents of the original dynamics;
    \item [(c1)] The model reconstructs the physically dominant negative Lyapunov exponents of the original dynamics;
    \item [(c2)] The reconstructed physically dominant Lyapunov exponents persist under the perturbation of hyperparameters.
\end{itemize}
When a constructed model predicts trajectories well, (c0) is usually satisfied, but (c1) and (c2) are not necessarily satisfied.
A model that does not meet at least (c1) or (c2) is called a {\bf non-ideal model}.
For an ideal model, we define Lyapunov exponents, which are the same as the physically dominant Lyapunov exponents of the original dynamics, as the {\bf actual Lyapunov exponents}.
We denote the positive and negative actual Lyapunov exponents as $\lambda_{\rm p}$ and $\lambda_{\rm n}$, respectively.
The number of the model variables is usually higher than the number of physically dominant Lyapunov exponents of the original dynamics. 
Therefore, the set of physically dominant Lyapunov exponents excludes some of the Lyapunov exponents of the model.
The exponents are called the {\bf spurious Lyapunov exponents}~\footnote{For non-ideal models, the Lyapunov exponents originating from the actual or spurious Lyapunov exponents of the ideal model are called the actual or spurious Lyapunov exponents, respectively.}.
We denote the $i$-th spurious Lyapunov exponent as $\lambda_{{\rm s}_i}$.
Figure~\ref{fig:lyap-img} provides a schematic diagram illustrating the various definitions of Lyapunov exponents.

Some works have examined the embedding methods concerning the reconstruction of attractors from time series data.
For example, some studies~\cite{takens_1981,sauer_1991a} report that an attractor of the original dynamics can be reconstructed as an embedding in the delay coordinate of a scalar observable variable of the original dynamics.
W. Ott and J.A. Yorke~\cite{ott2003} theoretically prove that the embedded attractor on the delay coordinate is realized on a low-dimensional manifold.
The Lyapunov vectors corresponding to the actual Lyapunov exponents\footnote{The main body defines the actual/spurious Lyapunov exponents only for data-driven models.
For a delay-embedded system, they can be defined similarly for data-driven models; the actual Lyapunov exponents are Lyapunov exponents in the original dynamics, and the others are the spurious Lyapunov exponents.} 
are on the tangent space of the low-dimensional manifold; those corresponding to the spurious Lyapunov exponents are not. 
The same geometric structure is expected in the data-driven ideal models.
Other studies~\cite{sauer1998, tempkin2007} theoretically prove that when the attractor of the two-dimensional system is embedded in the five-dimensional space, the reconstructed system has the spurious Lyapunov exponents $2\lambda_1, \lambda_1+\lambda_2, 2\lambda_2$,
where $\lambda_1$ and $\lambda_2$ are the actual Lyapunov exponents.

The main contribution of this paper is to reveal the geometric structure of ideal data-driven models and to show that the robustness of physically dominant Lyapunov exponents is a direct consequence of this structure.
For both discrete and continuous time dynamics, we attempt to identify an ideal RfR model that meets the conditions (c1) and (c2) in addition to (c0).
Using the Lyapunov vectors~\cite{ginelli_2007,saiki_2010}, we numerically verify that our model chaotic set (either a chaotic attractor or a chaotic saddle) is an embedding in the high-dimensional model space.
Lyapunov vectors corresponding to the actual Lyapunov exponents belong to the tangent space of the low-dimensional manifold on which the original dynamics is reproduced.
We demonstrate that the actual Lyapunov exponents persist robustly under changes in hyperparameters, which is crucial for distinguishing between ideal and non-ideal models, even when the actual Lyapunov exponents are unknown.
We investigate data-driven models constructed using the RfR method for two chaotic dynamics: the H\'enon map and the Lorenz system.

\section{Construction of a model: Radial function-based regression (RfR)}
\label{sec:method}

This section briefly explains the RfR method~\cite{tsutsumi22,tsutsumi23,tsutsumi24}.
We suppose the original system generating an observable time series is described as follows:
\begin{align}
\begin{split}    
    & \bm{x}_{n+1} = f(\bm{x}_n)~~(\rm{discrete-time}),\\
    & \frac{d\bm{x}}{dt} = f(\bm{x})~~~~~~~~(\rm{continuous-time}).
\end{split}
\end{align}
We observe some of the components of the variable $\bm{x}$, or more generally 
\begin{equation}
    \omega_1=g(\bm{x}). \label{eq:observable} 
\end{equation}
When the number of observable variables exceeds one, applying the joint RfR method~\cite{tsutsumi24}, a variant of the RfR method, is useful.

This paper focuses on the case where the number of observable variables is one.
We construct a $D$-dimensional model variable, called ${\bf X}$, to describe the dynamics behind the observed scalar time series.
When constructing a model, we introduce the delay coordinates of a scalar observable variable: 
${\bf X} := (\omega_1(t),\omega_1(t-\tau), \omega_1(t-2\tau),\ldots, \omega_1(t-(D-1)\tau))$, where $\tau$ is the delay time.
Theoretical studies~\cite{takens_1981, sauer_1991a} prove that the original attractor is reconstructed on the delay coordinated system by applying a proper delay time and a dimension.
This geometric structure should also be realized in the ideal model.
Note that the theories do not insist that the structure is realized for any data-driven models using delay coordinates.

We construct a model of a variable ${\mathbf X}$ as follows:
\begin{align}
\begin{split}
    &{\mathbf X}(t+1) = {\mathbf F}({\mathbf X}(t))~~(\rm{discrete-time}),\\
    &\frac{d{\mathbf X}(t)}{dt} = {\mathbf F}({\mathbf X}(t))~~~~~~~~(\rm{continuous-time}).
\end{split}\label{eq:mai_model}
\end{align}
We define the form of ${\mathbf F}({\mathbf X})(:=\{F_k(\mathbf{X})\}_{k=1,\ldots,D})$ as follows:
\begin{equation}
    F_k(\mathbf{X}) :=
    \tilde{\beta}_0^{(k)} + \sum_{d=1,\cdots,D} \tilde{\beta}_d^{(k)} X_d + \sum_{j=1,\cdots,J} \tilde{\beta}_{D+j}^{(k)}~\phi_j(\mathbf{X}),
\label{eq:gaussianpoly}
\end{equation}
where ${\tilde{\boldsymbol \beta}^{(k)}:=\{\tilde{\beta}_i^{(k)}\}_{i=0,\ldots,J+D}}$ is a set of estimated parameters determined from the time series,
$X_d$ is the $d$-th component of the model variable $\mathbf{X}$
and  
\begin{equation}
\phi_j(\mathbf{X}) = \exp\left(\frac{-\|\mathbf{X}-c_j\|^2}{\sigma^2}\right),\label{eq:gaussian}
\end{equation}
is the Gaussian radial basis function.
Here, $\|\cdot\|$ denotes the $l^2$ norm,
$c_j \in \mathbb{R} ^D$ is the coordinate of the $j$~th center point ($j=1,\ldots,J$), and $\sigma^2$ is the parameter that determines the deviation of $\phi_j$.
This RfR method distributes $c_j$ as lattice points with grid size $\delta_{\rm grid}$.
For a given integer $m$ explained below, 
we consider $c_j$ such that 
data points exist in the $(m-1)\delta_{\rm grid}$-neighborhood.
An increase in $\delta_{\rm grid}$ decreases the number of center points and the required computational resources.
Although non-uniform placement of the center points, for example through optimization, may improve efficiency, we adopt a uniform placement for simplicity.
For a given $\delta_{\rm grid}$, $\sigma^2$ is determined as follows:
\begin{align}
     \sigma^2 := \frac{((m-1) \delta_{\rm grid})^2}{- \log_{e} p},\label{eq:deviation}
\end{align}
where $m$ is the degree of the corresponding B-spline basis function and $p~(>0)$ is a small value.

\begin{table}[t!]
    \centering
    \caption{Explanations of variables and parameters used in the RfR method.}
    \scriptsize
    \begin{tabular}{|c|l|}
        \hline
        parameter & \multicolumn{1}{c|}{description} \\ \hline
        $\mathcal{D}$  &  dimension of the original dynamics  \\  \hline
        $\bm{x} (\in {\mathbb R}^\mathcal{D})$  & the variable of the original dynamics  \\  \hline
        $\bm{f}(\bm{x})$  & function defining the original system  \\  \hline
        $N$ & \makecell[l]{the number of the physically dominant Lyapunov exponents of\\ the original dynamics} \\  \hline
        $\omega_1 (\in \mathbb{R})$  &  scalar observable variable  \\  \hline
        $g(\bm{x})$  &  function generating an observable variable $\omega_1$ from $\bm{x}$  \\  \hline
        $D$  &  dimension of the model  \\  \hline
        ${\bf X} (\in {\mathbb R}^D)$  &  model variable  \\  \hline
        $\tau$&  delay time used for defining the model variable \\ \hline
        $\bf{F}(\bf{X})$& function defining the model system~\eqref{eq:mai_model} \\ \hline
        $v_k$& $k$-th component of a vector $\bm{v}$ \\ \hline
        $\tilde{\boldsymbol \beta}^{k}$& a set of  model parameters in \eqref{eq:gaussianpoly} estimated as \eqref{eq:solution}\\ \hline
        $\phi_j$& Gaussian radial basis function ~\eqref{eq:gaussian} \\ \hline
        $J$& the number of the Gaussian radial basis function ~\eqref{eq:gaussian} \\ \hline
        $c_j$& coordinate of a center point of $\phi_j$ \\ \hline
        $\sigma^2$ & a parameter that determines the deviation of $\phi_j$, defined in \eqref{eq:deviation} \\ \hline
        $\delta_{\rm grid}$ & grid size of lattice points used for distribution of $c_j$ \\ \hline
        $m$ & \makecell[l]{a parameter determining the radius of the neighborhood used\\ for the distribution of $c_j$ }\\ \hline
        $p$ & a small value determining $\sigma^2$ \\ \hline
        $N_T$ & the data size of the observed time series \\ \hline
        $n$ &  sampled data size for regression \\ \hline
        $t_i$ & time of the $i$-th sampled point \\ \hline
        $\frac{dX_k}{dt}|_{\rm est}$ & estimated numerical derivative\\ \hline
        $L_k(\bm{b})$ & \makecell[l]{optimization function \eqref{eq:L_k} to define the linear coefficients\\ of the ODE for $X_k$}\\ \hline
        $\bm{y}_k$ & numerically centered $\frac{dX_k}{dt}|_{\rm est}$ for linear regression \\ \hline
        $\eta_k$ & the standard deviation of the components of the $\bm{y}_k$\\ \hline
        $\alpha$ & regularization parameter on regression \\ \hline
        $A$ & design matrix for linear regression \\ \hline
    \end{tabular}
\label{tab:def-of-params}
\end{table}

The coefficients $\tilde{\beta}_i^{(k)}$ in \eqref{eq:gaussianpoly} are determined to predict a next step value (for a discrete-time system) or a derivative (for a continuous-time system).
Since we use the numerical derivatives to construct a model for a continuous-time system,
we apply the Taylor approximation to estimate numerical derivatives~$\frac{dX_k(t_i)}{dt}|_\mathrm{est}$.
See \cite{tsutsumi23} for more details about estimating derivatives.
Due to the limited computational resources,
we randomly choose samples of size $n$ among $N_T$ ($n\ll N_T$) for regression.
We denote the time of the $i$-th sampled point as $t_i~(i=1,\ldots,n)$.
The coefficients $\tilde{\boldsymbol \beta}^{(k)}$ are obtained as the minimizer $\bm{b}$ of the following function:
\begin{align}
    L_k(\bm{b}) = \frac{1}{2n} \|\bm{y}_k - A \bm{b} \|^2 + \frac{\alpha \eta_k}{2} \|\bm{b}\|^2.
    \label{eq:L_k}
\end{align}
Here,  
\begin{align*}
    \bm{y}_k := (z^{(k)}_1-\overline{z}^{(k)},z^{(k)}_2-\overline{z}^{(k)} \ldots, z^{(k)}_n-\overline{z}^{(k)})^\mathrm{T},
\end{align*}
where 
\begin{align*}
    z^{(k)}_i := 
    \begin{cases}
        X_k(t_i+1)~({\rm discrete-time}),\\
        \frac{dX_k(t_i)}{dt}|_\mathrm{est}~({\rm continuous-time}),
    \end{cases}
\end{align*}
and $\overline{z}^{(k)}$ is the mean of $\{z^{(k)}_i\}$,
and $\alpha$ is a parameter determining the strength of regularization, 
$\eta_k$ is a standard deviation of $\bm{y}_k$,
and $A$ is $n \times (D+J)$ matrix whose $i$-th row is centered
$(X_1(t_i), \ldots X_D(t_i), \phi_1(\bm{X}(t_i)), \ldots, \phi_J(\bm{X}(t_i)) )$ 
(see \eqref{eq:gaussianpoly}).
Regularization prevents overfitting, primarily by introducing the Gaussian radial basis functions.
The minimizer
of $L_k(\bm{b})$ is written as follows: 
\begin{equation} 
    \tilde{\boldsymbol \beta}^{(k)}=({A^\mathrm{T}} A + n \alpha \eta_k I )^{-1} {A^\mathrm{T}}  \bm{y}_{k}, \label{eq:solution}
\end{equation}
where $I$ is the identity matrix and $A^\mathrm{T}$ is the transpose of $A$.

The $\delta_{\rm grid}$ parameter should correspond to a scale of the variable $\omega_1$ in \eqref{eq:observable}; hence, $\omega_1$ is standardized in our modeling to avoid the adjustment.
We consider the following factors when selecting the parameters.
\begin{itemize}
    \item The model's dimension $D$ is selected to be higher than the expected attractor dimension. 
    \item The delay time $\tau$ is selected based on the decay of correlation of a variable $X_1(t)$. $\tau$ is chosen to meet that the correlation between $X_1(t)$ and $X_1(t-\tau)$ is around 0.5.
    \item Since the center points $c_j$ are distributed as lattice points, the settings of $\delta_{\rm grid}$ and $m$ determine the number of center points($J$). 
    In the settings shown in Table.~\ref{tab:def-of-params}, $J=1522$ for the H\'enon map and $J=12,968$ for the Lorenz system.
     \item The number of regression points ($n$) is not very sensitive and is fixed at 50,000.
\end{itemize}

In data-driven modeling from a scalar time series, it is generally necessary to use a sufficiently high embedding dimension in order to reconstruct the original dynamics as an embedding. 
According to the theory of delay-coordinate embedding (e.g., Takens' theorem~\cite{takens_1981}), the required dimension can exceed twice the dimension of the original system.
In contrast, models constructed in lower dimensions may reproduce trajectories or invariant measures reasonably well, but often fail to reconstruct the full dynamical structure, in particular the negative Lyapunov exponents~\cite{tsutsumi22}. 
Therefore, in this study, we consider models with embedding dimensions larger than the intrinsic dimension of the original system.

Among the hyperparameters, the regularization parameter \( \alpha \) plays the most important role in determining the quality of the model and is systematically investigated in this study. 
In contrast, the other parameters are selected based on standard heuristics used in delay-coordinate modeling and are not highly sensitive within reasonable ranges.
We confirmed that the conclusions of this study remain unchanged for such parameter choices.

Integrating the model \eqref{eq:gaussianpoly} and \eqref{eq:mai_model} over time allows us to generate a trajectory of the constructed data-driven model. 
When a model trajectory is realized on a chaotic set of the model that is not an attractor, it cannot generate a long-term trajectory.
Therefore, we employ the Stagger-and-Step method~\cite{sweet_2001c} to generate a long-term trajectory of the constructed model.

\section{Construction of ideal models}

This section constructs ideal models that capture the physically dominant Lyapunov exponents for well-known discrete and continuous time chaotic dynamics: the H\'enon map and the Lorenz system.
Our previous study~\cite{tsutsumi22} reported that the constructed model, using a set of hyperparameters, does not reconstruct the negative Lyapunov exponent, whereas it does reconstruct the non-negative ones.
Even if the negative Lyapunov exponents are not reconstructed in the constructed model, the model can still approximate a short-term trajectory and statistical quantities generated from a long-term trajectory. 
When the hyperparameters are chosen appropriately, even the negative Lyapunov exponents should be reconstructed in the constructed model.

\subsection{Sample dynamics}

\begin{table}[t!]
\scriptsize
    \centering
    \caption{
        Hyperparameters used for the modeling of each example's dynamics.
    }
    \begin{tabular}{|c||c|c|c|c|}
    \hline
         & \multicolumn{2}{c|}{H\'enon} &\multicolumn{2}{c|}{Lorenz}  \\ \cline{2-5}
         & Model A & Model B & Model A & Model B  \\ \hline
         $D$& \multicolumn{2}{c|}{4} & \multicolumn{2}{c|}{6}   \\ \hline
         $\tau$& \multicolumn{2}{c|}{1}& \multicolumn{2}{c|}{0.03}  \\ \hline
         $\delta_{\rm grid}$& \multicolumn{2}{c|}{0.50}& \multicolumn{2}{c|}{0.50}  \\ \hline
         $p$&\multicolumn{2}{c|}{0.001} & \multicolumn{2}{c|}{0.0005}\\ \hline
         $m$&\multicolumn{2}{c|}{3} & \multicolumn{2}{c|}{3}\\ \hline
         $N_T$&\multicolumn{2}{c|}{1,000,000} & \multicolumn{2}{c|}{1,000,000}\\ \hline
         $n$&\multicolumn{2}{c|}{50,000} & \multicolumn{2}{c|}{50,000}\\ \hline
         $\log_{10} \alpha$& -11.0 & -5.5 & -13.0 & -9.0 \\ \hline
    \end{tabular}
    \label{tab:used-params}
\end{table}

As a typical example of the discrete-time chaotic dynamics, this study deals with the H\'enon map: 
\begin{align*}
    &x_{n+1} = 1 - 1.4 x_n^2 + y_n,\\
    &y_{n+1} = 0.3 x_n.
\end{align*}
The Lyapunov exponents are $ 0.42$ and $ -1.62$, each belonging to the physically dominant Lyapunov exponents.
Furthermore, as a typical example of the continuous-time chaotic dynamics, we deal with the Lorenz system:
\begin{align*}
    &\frac{dx}{dt} = 10 ( y - x ),\\
    &\frac{dy}{dt} = 28 x - y -xz,\\
    &\frac{dz}{dt} = xy - \frac{8}{3}z.
\end{align*}
The Lyapunov exponents are $0.91, 0.00, -14.57$; all are the physically dominant Lyapunov exponents. 
To generate time series used in the training and evaluation, the fourth-order Runge-Kutta is applied with $\Delta t=0.01$.
In each example, we assume the observable variable is only the first variable ($x_n$ and $x$, respectively).

\subsection{Results for the H\'enon map}
\label{sec:example-henon}

\begin{table}[t!]
    \centering
    \caption{
        Lyapunov exponents of Model A (ideal model) and Model B (non-ideal model) for the H\'enon map.
    }
    \begin{tabular}{|c|c|c|c|c|}
        \hline
         & Positive &  Negative &  \multicolumn{2}{c|}{Spurious} \\ \cline{4-5} 
         & $\lambda_{p}$ &  $\lambda_{n}$&  $\lambda_{{\rm s}_1}$ & $\lambda_{{\rm s}_2}$ \\ \hline
        H\'enon map    & 0.419 & -1.622 & -- & -- \\\hline
        Model A & 0.417 & -1.608 & -0.325 & -0.769 \\ \hline
        Model B & 0.410 & -1.518 & -0.498 & -0.746 \\ \hline
    \end{tabular}
    \label{tab:lyap-compare_henon}
\end{table}

\begin{figure}[t]
    \begin{center}
        \includegraphics[width=0.90\linewidth]{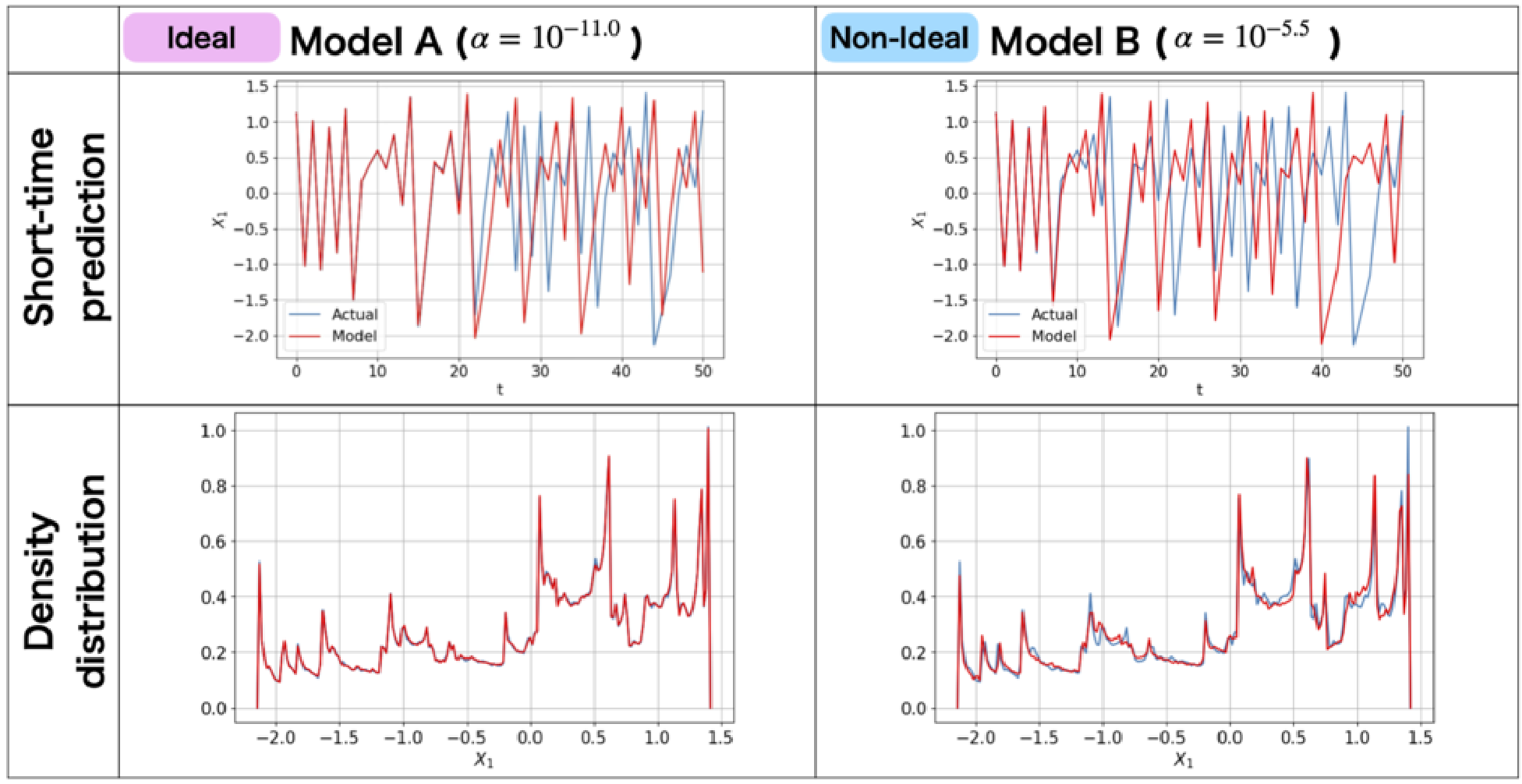}
    \end{center}
    \caption{
    {\bf Trajectories of the constructed models~(Model A and Model B) for the H\'enon map.} 
    The red lines represent the results of the constructed models, whereas the blue lines are for the actual.
    Model A is an ideal model and Model B is not~(See Table~\ref{tab:lyap-compare_henon}).
    For not only Model A but also Model B, a short-term trajectory and density distribution created from a long-term trajectory are reconstructed well.
    The reconstruction of trajectories is more accurate when all of the physically dominant Lyapunov exponents are successfully reproduced.
    }
    \label{fig:traj-henon}
\end{figure}

\begin{figure}
    \begin{center}
        \includegraphics[width=\linewidth]{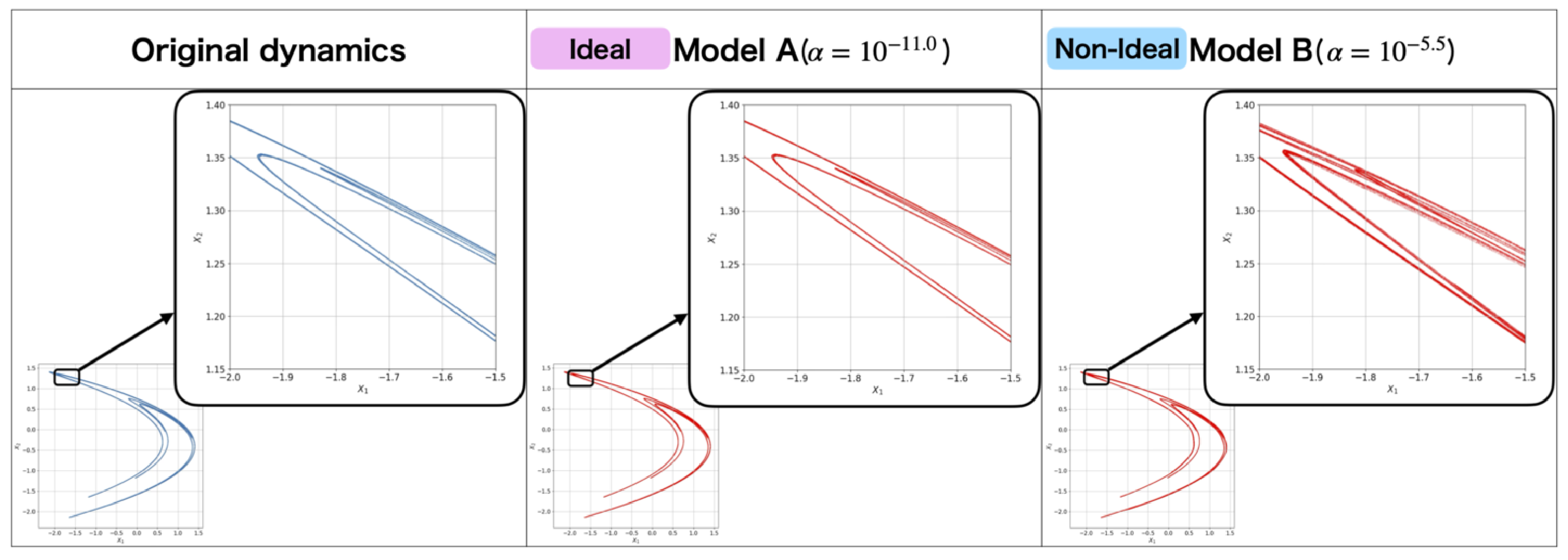}
    \end{center}
    \caption{
    {\bf Comparison of model attractors projected on the 2-dimensional space.} 
    The model attractors created by long-term trajectories of two different models are projected on the ($X_1, X_2$)-plane.
    The attractor of the ideal model is similar to that of the original dynamics, while that of the non-ideal model is blurred compared to that of the original dynamics.
    The fact that the actual negative Lyapunov exponent of the non-ideal model is larger than that of the original dynamics results in a larger Lyapunov dimension of the reconstructed chaotic invariant set of the non-ideal model, which causes a slightly weaker predictability of trajectories shown in Fig.~\ref{fig:traj-henon}.
    See Table~\ref {tab:lyap-compare_henon} for the Lyapunov exponents. 
    }
    \label{fig:2d-plot-henon}
\end{figure}

We construct four-dimensional data-driven models using the RfR method with two different regularization parameters for the H\'enon map: 
$\alpha=10^{-11}$ (Model A) and $\alpha=10^{-5.5}$ (Model B). 
Table~\ref{tab:used-params} represents a set of other hyperparameters used for the models. 
Table~\ref{tab:lyap-compare_henon} indicates the Lyapunov exponents of the constructed models. 
Here, we estimate the Lyapunov exponents using the Jacobian matrix of the model and the QR decomposition~\cite{ginelli_2007}, using trajectories of length 1,000,000.
The number of the models' variables is higher than that of the original system; therefore, some spurious Lyapunov exponents exist. 
Model A reconstructs the original system's physically dominant Lyapunov exponents, and Model B does not.
This means that Model A is an ideal model and Model B is not.
How to construct an ideal model, which is the same as how to choose hyperparameters, is discussed in the following section.

Figure~\ref{fig:traj-henon} shows a short-term trajectory and density distribution computed from a long-term trajectory.
Each model predicts a short-term trajectory well, and the invariant density distribution is similar to that of the original dynamics.
The constructed model can describe a trajectory well, even when it is not an ideal model.
The reconstruction of trajectories is more accurate when all of the physically dominant Lyapunov exponents are successfully reproduced.
Figure~\ref{fig:2d-plot-henon} compares attractors of an ideal model and a non-ideal model. 
The negative Lyapunov exponent of the non-ideal model is larger than that of the original dynamics; therefore, the attractor of the non-ideal model looks blurred, which decreases the trajectory accuracy.

\subsection{Results for the Lorenz system}
\label{sec:example-lorenz}

\begin{table}[t!]
    \centering
    \caption{
        Lyapunov exponents of Model A (ideal model) and Model B (non-ideal model) for the Lorenz system.
    }
    \resizebox{\linewidth}{!}{
    \begin{tabular}{|c|c|c|c|c|c|c|}
        \hline
         & Positive &  Neutral &  Negative&  \multicolumn{3}{c|}{Spurious} \\ \cline{5-7} 
         & $\lambda_{p}$ &   &  $\lambda_{n}$&  $\lambda_{{\rm s}_1}$ & $\lambda_{{\rm s}_2}$ & $\lambda_{{\rm s}_3}$ \\ \hline
        Lorenz system & 0.906 &  0.000 & -14.572 & -- & -- & -- \\ \hline
        Model A& 0.906 & -0.011 & -14.517 & 0.101 & -0.233 & -4.028 \\ \hline
        Model B& 0.896 & -0.012 & - 9.298 & -0.033 & -0.168 & -1.176 \\ \hline
    \end{tabular}
    }
    \label{tab:lyap-compare_lorenz}
\end{table}

\begin{figure}[t]
    \begin{center}
        \includegraphics[width=0.9\linewidth]{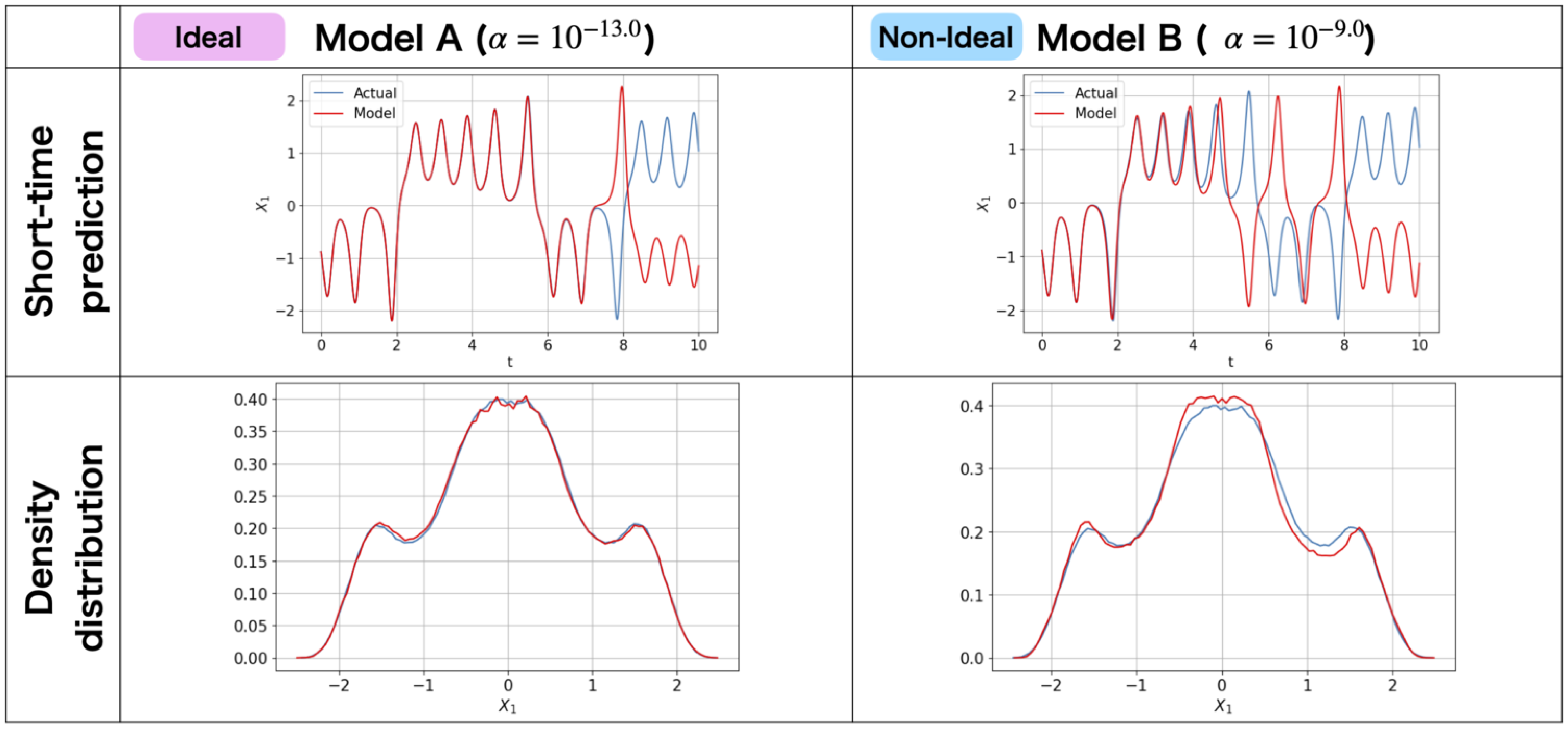}
    \end{center}
    \caption{
    {\bf Trajectories of the constructed models~(Model A and Model B) for the Lorenz system.} 
    This figure is the H\'enon map's version of Fig.~\ref{fig:traj-henon}.
    As is the case with the H\'enon map, the reconstruction of trajectories is more accurate when all of the physically dominant Lyapunov exponents are successfully reproduced.
    }
    \label{fig:traj-lorenz}
\end{figure}

We construct six-dimensional data-driven models using the RfR method with two different regularization parameters for the Lorenz system: $\alpha=10^{-13.0}$ (Model A) and $\alpha=10^{-9.0}$ (Model B). 
The other settings are in Table~\ref{tab:used-params}.
Table~\ref{tab:lyap-compare_lorenz} presents the Lyapunov exponents of the constructed models.
Model A reconstructs all of the physically dominant Lyapunov exponents; Model B does not reconstruct the negative Lyapunov exponent of the original system.
We can conclude that Model A is the ideal model and Model B is not.
Figure~\ref{fig:traj-lorenz} presents the successful reconstruction of both short- and long-term trajectories, showing that Model A has slightly higher accuracy than Model B.

\section{Geometric structure of the constructed model}
\label{sec:geometric_structure}

\begin{figure}
    \begin{center}
        \includegraphics[width=\linewidth]{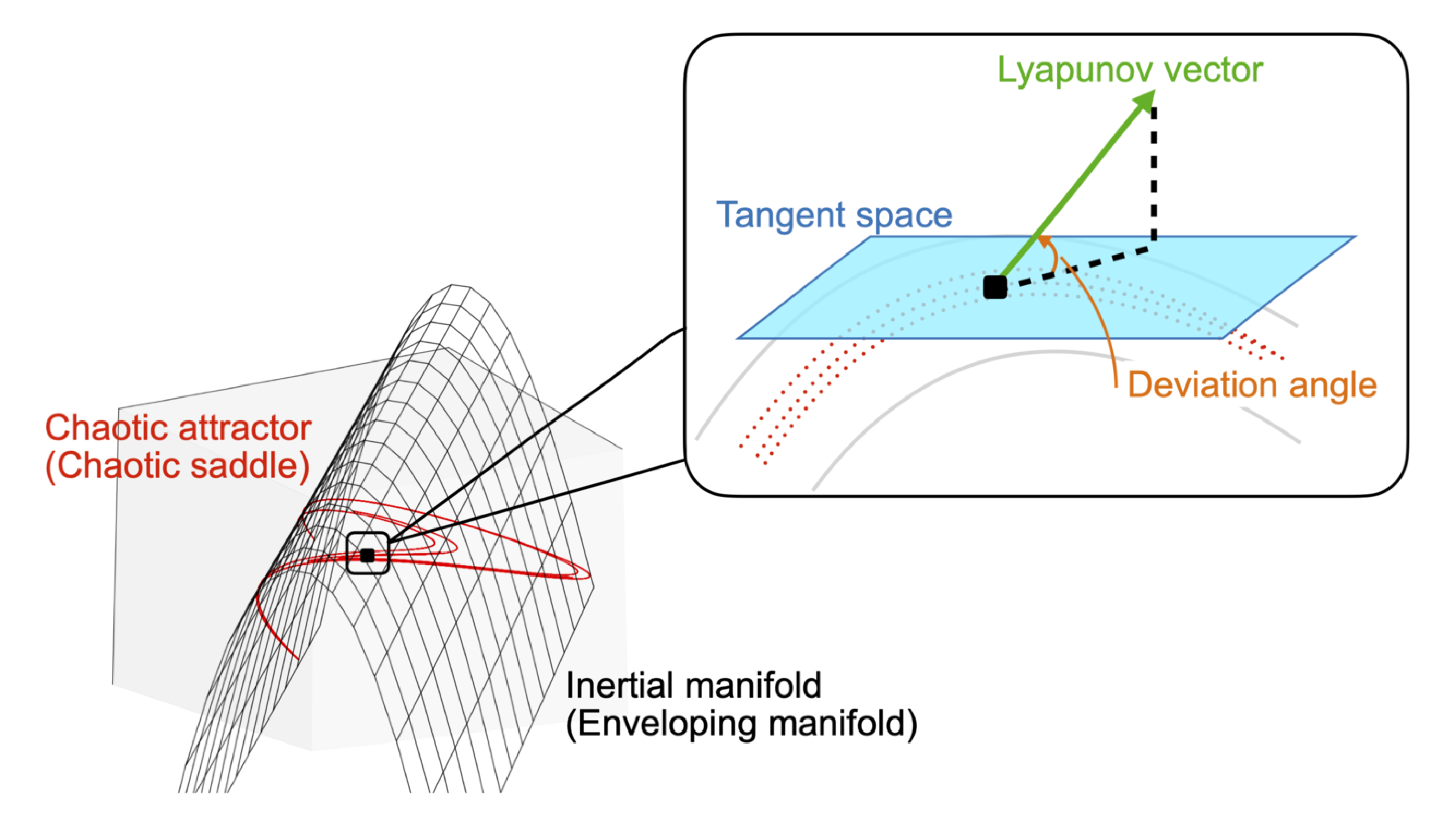}
    \end{center}
    \caption{
     {\bf Schematic picture of the geometric structure.}
     A model's chaotic attractor (or the chaotic saddle) is realized on a low-dimensional internal manifold (or the enveloping manifold).
     To show that the invariant set is realized on the manifold as an embedding, we calculate an angle between the Lyapunov vector and the tangent space of the low-dimensional manifold; the deviation angle.
    }
    \label{fig:img-deviation-angle}
\end{figure}

Some previous papers~\cite{takens_1981,sauer_1991a} insist that under some assumptions the attractor of dynamics in the original phase space can be embedded in a scalar observation's high-dimensional delay coordinate space of a scalar observation.
W. Ott and J.A. Yorke~\cite{ott2003} insist that the low-dimensional manifold exists in the delay coordinate space, and the invariant set corresponding to the attractor in the original phase space is realized on the manifold.
Figure~\ref{fig:img-deviation-angle} presents the schematic picture of the structure.
In some data-driven dynamical modeling~\cite{berry2023,hart2024,berry2025}, a similar embedding structure is expected to be realized and a key to reconstructing the original dynamics.
In this section, we investigate the geometric structure of the constructed ideal model using the RfR method and conclude that the original dynamics is realized as an embedding in the ideal model.
This structure contributes to the high accuracy of the trajectory prediction.

To analyze the geometric structure of the model attractor, we focus on the Lyapunov vectors of the model.
The Lyapunov vector is a locally linearized stable/unstable manifold of the dynamics~\cite{saiki_2010,ginelli_2007,kobayashi_2014}.
The Lyapunov vectors corresponding to the actual Lyapunov exponents are called the actual Lyapunov vectors; those corresponding to the spurious Lyapunov exponents are the spurious Lyapunov vectors.
A previous study~\cite{ott2003} theoretically proves that the actual Lyapunov vectors belong to the tangent space of the manifold on which the original dynamics is reconstructed, and the spurious Lyapunov vectors do not.

An inertial manifold is a finite-dimensional, smooth, invariant manifold that contains the global attractor and attracts all solutions exponentially quickly.
\footnote{No inertial manifold exists if the model dynamics is realized as a chaotic saddle; however, even in that case, we can do a similar analysis using the {\bf enveloping manifold}~\cite{ott2003}.}
The inertial manifold is a low-dimensional manifold where the original dynamics are reconstructed.
In this study, we do not explicitly represent the inertial manifold itself; instead, we focus on its local tangent structure estimated from sample points along the attractor. 
Specifically, for each point on the model trajectory, we use nearby trajectory points to approximate a low-dimensional linear subspace via singular value decomposition, which is interpreted as the local tangent space of the inertial manifold.
We measure the angle between the Lyapunov vector at a model trajectory point and the tangent space of the model's inertial manifold.
The angle indicates how the Lyapunov vector deviates from the tangent space; hence, we call it the {\bf deviation angle}.
Figure~\ref{fig:img-deviation-angle} presents a schematic picture of the deviation angle.
We write the deviation angle for the Lyapunov vector of the actual positive Lyapunov exponent as $\theta_{\rm p}$.
That for the Lyapunov vector of the actual negative Lyapunov exponent is $\theta_{\rm n}$, and that for the Lyapunov vector of the $i$-th spurious Lyapunov exponent is $\theta_{{\rm s}_i}$. 
A slight deviation angle indicates that the corresponding Lyapunov vector belongs to the tangent space of the inertial manifold, suggesting that the original dynamics' attractor is reconstructed in the model space as an embedding.

\subsection{Results for the H\'enon map}
\label{sec:geo-results-henon}

\begin{figure}
    \centering
    \includegraphics[width=0.9\linewidth]{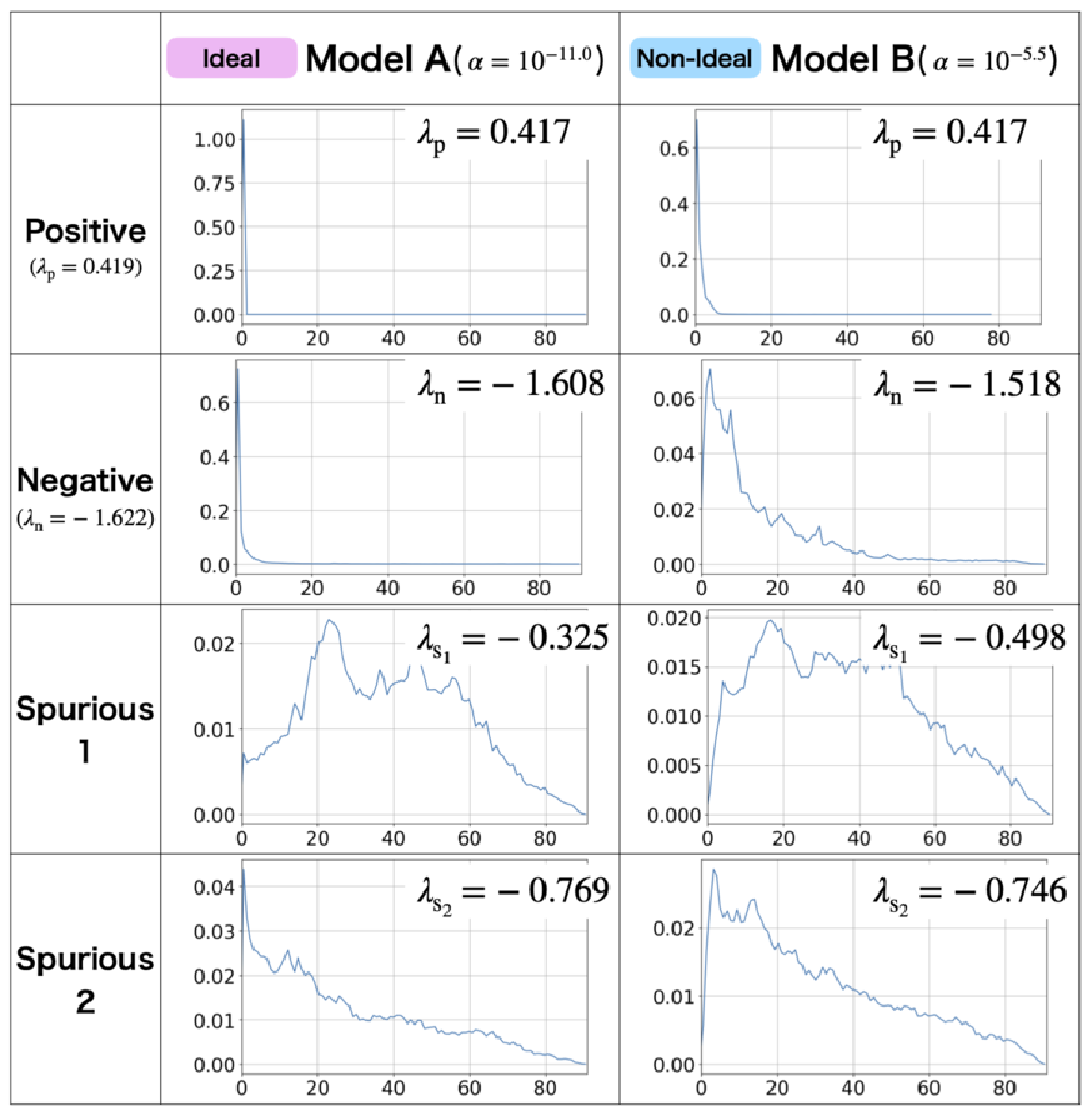}
    \caption{
    {\bf Distribution of deviation angles (degree) for the models of the H\'enon map.}
    For Model A ($\alpha = 10^{-11}$), the deviation angles for the actual positive and negative Lyapunov vectors~($\theta_{\rm p}$ and $\theta_{\rm n}$) are nearly zero at almost all trajectory points, whereas those for the spurious Lyapunov vectors~($\theta_{{\rm s}_i}$) show various values. 
    For Model B ($\alpha = 10^{-5.5}$), the peak of the density distribution of $\theta_{\rm p}$ is near zero, whereas $\theta_{\rm n}$ shows a wide range of values.
    The results imply that the dynamics is realized in Model A as embedding but not in Model B.
    }
    \label{fig:henon-lyap-vecs}
\end{figure}

We analyze the constructed models (Model A and Model B) for the H\'enon map.
Before calculating the deviation angles, the dimension of the inertial manifold needs to be estimated.
As the box-counting dimension of a model attractor of the H\'enon map is lower than two, the dimension of the inertial manifold is two.
We calculate the deviation angle 
at each sample point of a model of the H\'enon map.
For the calculation, we estimate the two-dimensional tangent space at a sample point from the time series of a trajectory as follows. 
We take trajectory points in the neighborhood of each sample point and define a two-dimensional linear space where the selected points lie using singular value decomposition~\cite{zu1997}.

Figure~\ref {fig:henon-lyap-vecs} shows the distributions of the deviation angles for models of the H\'enon map with multiple regularization parameters.
The concentration of the distribution at zero means that the Lyapunov vectors are on the two-dimensional tangent space of the inertial manifold.
When the original dynamics is reconstructed as an embedding in the model space,
the actual Lyapunov vectors should be on the tangent space and the spurious Lyapunov vectors should not.
Concerning Model A, an ideal model obtained using $\alpha = 10^{-11}$, the actual Lyapunov vectors are on the tangent space of the model's inertial manifold.
The deviation angles for the spurious Lyapunov vectors~($\theta_{{\rm s}_i} $) take various values.
Concerning Model B, the non-ideal model obtained using $\alpha=10^{-5.5}$, the actual negative Lyapunov exponent differs from the actual value; hence, the deviation angle for the actual negative Lyapunov vector~($\theta_{\rm n}$) takes various values.
These results suggest that for the ideal model~(Model A), the original dynamics is reconstructed as an embedding in the model.
For the non-ideal model~(Model B), the dynamics is not.

\subsection{Results for the Lorenz system}
\label{sec:geo-results-lorenz}

\begin{figure}[!t]
    \centering
    \includegraphics[width=0.9\linewidth]{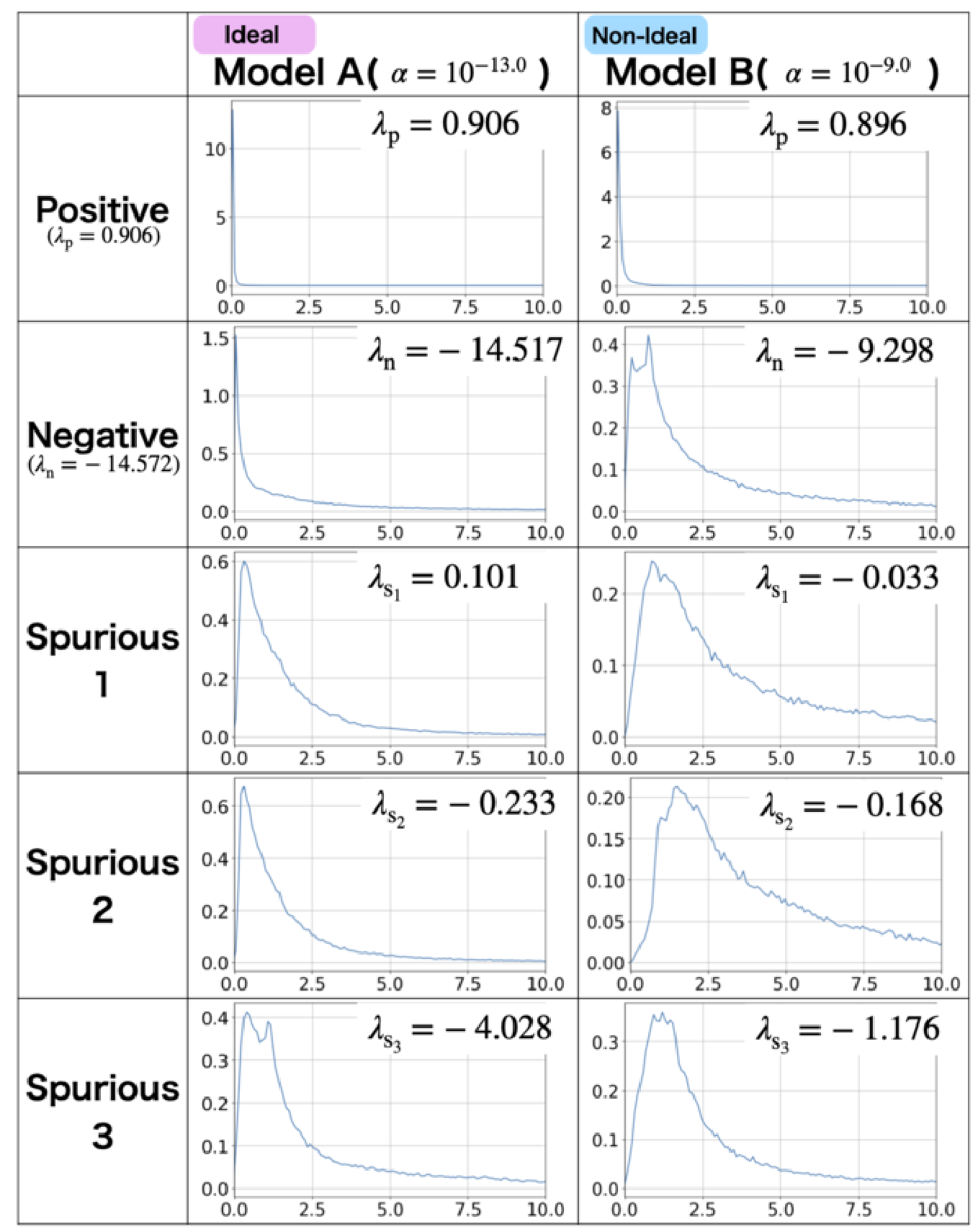}
    \caption{
    {\bf Distribution of deviation angles (degree) for the models of the Lorenz system.}
    For Model A ($\alpha = 10^{-13}$), the deviation angles for the adjusted actual positive and negative Lyapunov vectors~($\theta_{\rm p}$ and $\theta_{\rm n}$) are nearly zero at almost all trajectory points, whereas those for the adjusted spurious Lyapunov vectors~($\theta_{{\rm s}_i}$) show various angle values. 
    For Model B ($\alpha = 10^{-9}$), the peak of the density distribution of $\theta_{\rm p}$ is near zero, 
    whereas $\theta_{\rm n}$ shows a wide range of values. 
    The results imply that the dynamics is realized in Model A as embedding but not in Model B, which is the same as the case of the H\'enon map.
    }
    \label{fig:lorenz-lyap-vecs}
\end{figure}

We apply a similar analysis for the Lorenz system using the constructed models (Models A and B).
The box-counting dimension of the model attractor is lower than three, and the dimension of the inertial manifold of the model attractor is three.
As the models are continuous-time systems, we consider the orbit direction in the calculation.
We define a vector obtained by removing the orbit direction from the Lyapunov vector as the {\bf adjusted Lyapunov vector} and a two-dimensional linear space obtained by removing the orbit direction from the tangent space, which is three-dimensional, as the {\bf adjusted tangent space}.
We use the angles between the adjusted Lyapunov vector and the adjusted tangent space as the deviation angles for the continuous-time system.
The orbit direction is directed toward the Lyapunov vector of the neutral Lyapunov exponent; therefore, the deviation angle for the neutral Lyapunov exponent cannot be defined.

The Lorenz system produces results similar to those of the H\'enon map.
Figure~\ref{fig:lorenz-lyap-vecs} represents the distribution of the deviation angles for models of the Lorenz system.
The deviation angles for the spurious adjusted Lyapunov vectors~($\theta_{{\rm s}_i}$) take various values. 
The deviation angles for the adjusted actual positive and negative Lyapunov vectors~($\theta_{\rm p}$ and $\theta_{\rm n}$) take $0$ for Model A, which is an ideal model.
For the non-ideal model~(Model B), $\theta_{\rm n}$ takes various values.
Therefore, we can conclude that the original dynamics is reconstructed on Model A as an embedding, not Model B.

\section{Finding a set of hyperparameters for the ideal model}
\label{sec:how2construc}

In practical situations, the Lyapunov exponents of the original system are unavailable, and it is not obvious how to determine whether a constructed model is ideal.
In this section, we propose a criterion for identifying ideal models based solely on properties of the constructed model.
The embedding structure is considered a characteristic of a successfully constructed model.
Similarly, the fixed points of the actual system should also be reproduced in the successful model.
Previous paper~\cite{tsutsumi22} reports that fixed points exist robustly for hyperparameters in the constructed models, which predict trajectories well. 
The models have some spurious fixed points, and they are highly dependent on changes in hyperparameters.
It is thought that the structures existing in the original dynamics would be preserved and persist under the change in hyperparameters, whereas structures that do not exist in the original dynamics would depend on the hyperparameters.
In this sense, the actual Lyapunov exponents should exist robustly, but the spurious exponents should highly depend on the change of hyperparameters.
This dependency is crucial for distinguishing actual exponents from spurious ones and selecting a set of hyperparameters for the ideal model.
To verify this hypothesis, we construct models with various values of hyperparameters and calculate Lyapunov exponents.
We confirmed the robustness of the physically dominant Lyapunov exponents and the sensitivity of spurious exponents across embedding dimensions (see Supplemental Material).

\subsection{Results for the H\'enon map}

\begin{figure}[t]
    \begin{center}
        \includegraphics[width=0.90\linewidth]{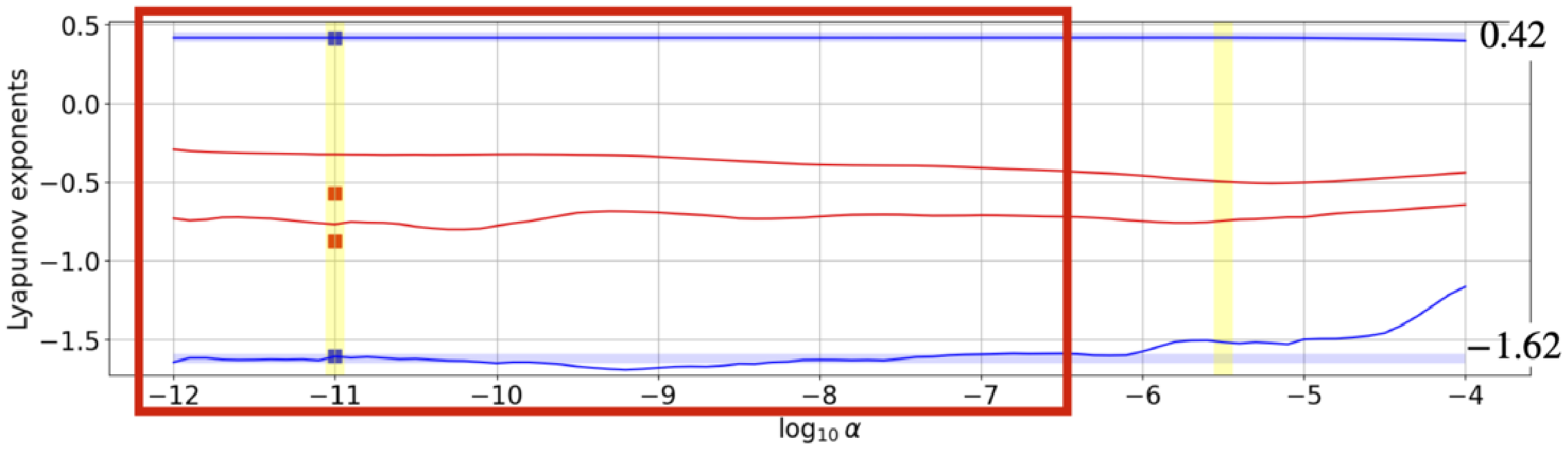}
    \end{center}
    \caption{
    {\bf Lyapunov exponents concerning the regularization parameter for the H\'enon map.} 
    The Lyapunov exponents are calculated for each model with a four-dimensional delay coordinate.
    The horizontal axis represents the regularization parameter $\alpha$ on a logarithmic scale, 
    while the vertical axis is the Lyapunov exponent.
    Vertical yellow lines correspond to $\log_{10}\alpha$ used for Model A and B in Table~\ref{tab:lyap-compare_henon}.
    Around $\log_{10}\alpha=-11.0$, the model's actual Lyapunov exponents take similar values to the physically dominant Lyapunov exponents (purple), which indicates that Model A is an ideal model.
    The first and fourth Lyapunov exponents (blue) remain relatively stable around the physically dominant Lyapunov exponents when $\log_{10} \alpha \in [-12,-6.5]$~(surrounded by a red box).
    Four points at $\log_{10} \alpha =-11.0$ represent the Lyapunov exponents of another model using a grid size $\delta_{\rm grid}=1.00$.
    The second and third Lyapunov exponents (red), which are spurious, can vary depending on the hyperparameters. 
    }
    \label{fig:lyap-all-henon}
\end{figure}

We construct models using various regularization parameters for the H\'enon map, and calculate the Lyapunov exponents for each model.
Figure~\ref{fig:lyap-all-henon} shows how much the Lyapunov exponents of constructed models depend on the regularization parameter.
Around the regularization parameter used in Model A, which is the ideal model, the physically dominant Lyapunov exponents are robustly reconstructed in the models.
Compared to the robust reconstruction of the actual Lyapunov exponents, spurious Lyapunov exponents depend on the regularization parameter.
In Fig.~\ref{fig:lyap-all-henon}, the Lyapunov exponents of the model using different grid size~$\delta_{\rm grid}$ with $\alpha = 10^{-11}$ are shown as red and blue points.
The spurious Lyapunov exponents depend on the grid size.
We can conclude that the Lyapunov exponent that robustly exists within a particular range is the actual one, and models using a set of hyperparameters within this range are ideal models.

\subsection{Results for the Lorenz system}

\begin{figure}[t]
    \begin{center}
        \includegraphics[width=0.90\linewidth]{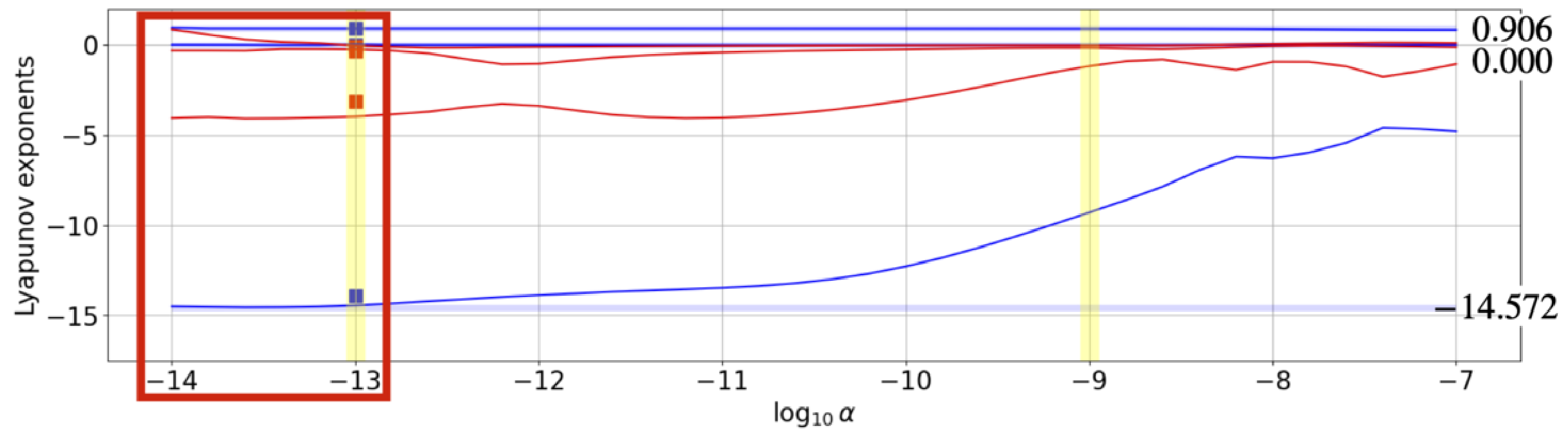}
    \end{center}
    \caption{
    {\bf Lyapunov exponents concerning the regularization parameter for the Lorenz system.}
    This figure is the Lorenz system's version of Fig.~\ref{fig:lyap-all-henon}.
    The model's actual Lyapunov exponents take similar values to the physically dominant Lyapunov exponents (purple) around $\log_{10}\alpha=-13.0$, meaning that Model A is an ideal model.
    The first and sixth Lyapunov exponents (blue) remain relatively stable around the physically dominant Lyapunov exponents when $\log_{10} \alpha \in [-14,-13]$~(surrounded by a red box).
    Six points at $\log_{10} \alpha =-13.0$ represent the Lyapunov exponents of another model using a grid size $\delta_{\rm grid}=0.40$.
    The other Lyapunov exponents (red), which are spurious, can vary depending on the hyperparameters. 
    }
    \label{fig:lyap-all-lorenz}
\end{figure}

We also analyze the relation between the actual/spurious Lyapunov exponents and values of the hyperparameters for the Lorenz system.
Figure~\ref{fig:lyap-all-lorenz} represents the extent to which Lyapunov exponents of constructed models depend on the regularization parameter.
Similar to the case of the H\'enon map, the original system's physically dominant Lyapunov exponents robustly exist around the regularization parameter used for the ideal model (Model A).
Figure~\ref{fig:lyap-all-lorenz} represents the Lyapunov exponents of the model using the different grid sizes as red and blue points.
The spurious exponents highly depend on the grid size.
The hyperparameter dependency can be a key to distinguishing between the actual and spurious Lyapunov exponents. 
When we take a set of hyperparameters from the robust range, a constructed model is an ideal.
Therefore, the robustness of Lyapunov exponents with respect to hyperparameters provides a practical way to identify ideal models even when the true system is unknown.

\section{Discussion}

When constructing a data-driven model, reconstructing local unstable manifolds that are characterized by non-negative Lyapunov exponents is essential for time series prediction.
To insist that the constructed model completely reproduces the original dynamics,
the model must reconstruct the fine structure, such as the fractal structure, in a stable direction.
Therefore, reconstructing the physically dominant Lyapunov exponents, the number of which corresponds to the Lyapunov dimension of the original attractor, is essential.
Note that the physically dominant Lyapunov exponents include some of the negative Lyapunov exponents for a chaotic system.
A model that reproduces all of the physically dominant Lyapunov exponents is called an ideal model.

Predictability of the short-term trajectories does not necessarily imply that the constructed model is an ideal model~\cite{tsutsumi22}.
For the ideal model, the reconstructed physically dominant Lyapunov exponents robustly exist under changes in the model's hyperparameters.
When the system dimension of the model is larger than that of the original system, there are some spurious Lyapunov exponents, which sensitively depend on hyperparameters.
These dependencies are a key to distinguishing the ideal model from the non-ideal model.
We emphasize that spurious Lyapunov exponents are not intrinsic properties of the original dynamics and are not used here as direct measures of model accuracy. 
Instead, their sensitivity to hyperparameters is used to distinguish them from the physically dominant Lyapunov exponents, which remain robust.

For the ideal model, the attractor of the original dynamics is found to be reconstructed in the model space as an embedding.
This outcome also means that a low-dimensional manifold, such as an inertial manifold or an enveloping manifold, exists in the model space, and the original dynamics is reproduced on the manifold.
This finding is numerically verified by confirming whether the Lyapunov vectors of the actual Lyapunov exponents belong to the tangent space of the manifold on which the original dynamics are reconstructed, and those of the spurious Lyapunov exponents do not.
The Lyapunov vectors corresponding to the actual Lyapunov exponents are constrained to lie on the tangent space with perturbations of hyperparameters; thus, the Lyapunov exponents exist robustly.
This behavior is why the actual Lyapunov exponents are independent of hyperparameters, while spurious ones depend on hyperparameters.

In this study, we construct ideal models for a two-dimensional map and a three-dimensional ordinary differential equation using delay coordinates.
Even when the original dynamics are high-dimensional or infinite-dimensional, such as delay differential equations or partial differential equations, the RfR method can construct models that describe the dynamics reasonably well~\cite{tsutsumi23}, and it is expected that ideal models may also be constructed in such cases.
However, when the Lyapunov dimension of the original dynamics is large, the computational cost can become prohibitive, since the number of center points increases exponentially with the Lyapunov dimension. 
Therefore, while the proposed approach is conceptually applicable to such systems, its practical implementation may be limited by computational resources.

In some cases, it may be possible to construct ideal models by incorporating non-delay features, such as past moving averages or standard deviations, to form high-dimensional coordinates. 
Since the proposed method relies on the geometric structure of the reconstructed dynamics, rather than the specific choice of coordinates, it is expected to remain applicable when such features are used, provided they capture relevant temporal and spatial scales.
Although models that describe the original dynamics well can be constructed from noisy time series~\cite{tsutsumi23}, the reconstruction of fine structures, in particular those associated with negative Lyapunov exponents, is sensitive to noise. 
While moderate noise levels may still allow for meaningful reconstruction, large noise amplitudes may destroy the underlying fractal structure, making it difficult to construct an ideal model.

The embedding structure is considered a key component in other data-driven modeling methods.
Some papers~\cite{berry2023, berry2025} theoretically verify that the embedding is realized in a constructed model, under some ideal assumptions.
Using our approach, which we apply to the RfR method, the numerical verification of the embedding structure will be obtained for models constructed using those methods without ideal assumptions.
Additionally, the approach will contribute to the hyperparameter tuning of other modeling methods, resulting in the improvement of the performance.

\section*{Acknowledgements}
YS was supported by the JSPS KAKENHI Grant No.25H01469.
KN was supported by the JSPS KAKENHI Grant No.22K17965.
The computation was carried out using the JHPCN (jh250021) 
and the Collaborative Research Program for Young $\cdot$ Women Scientists of ACCMS and IIMC, Kyoto University. 
The authors would like to thank the reviewers for their comments and criticisms to improve the manuscript.


\section*{References}
\bibliography{cas-refs.bib}

\clearpage
\onecolumngrid

\section*{Supplementary material: Additional results for different embedding dimensions}

\begin{figure}[h]
    \centering
    \includegraphics[width=0.45\linewidth]{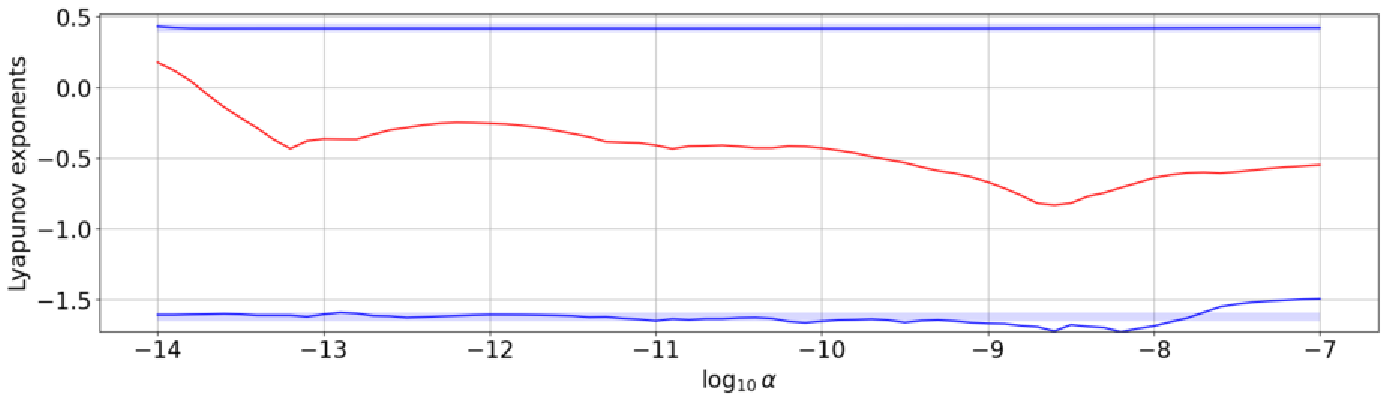}
    \includegraphics[width=0.45\linewidth]{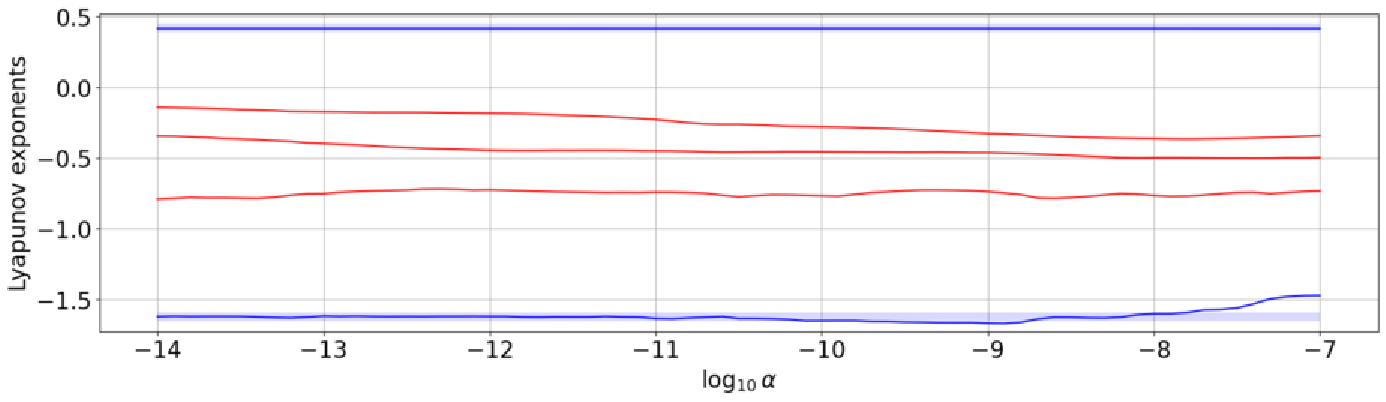}
    \caption{
    {\bf Lyapunov spectra of RfR models for the H\'enon map with different embedding dimensions.}
    {\it Left:} $D=3$. 
    {\it Right:} $D=5$.
    In both cases, the physically dominant Lyapunov exponents remain robust with respect to the hyperparameter, while the spurious exponents exhibit strong sensitivity. 
    This structure is consistent with the results shown in the main text.
}
    \label{fig:add-results-henon}
\end{figure}

\begin{figure}[h]
    \centering
    \includegraphics[width=0.45\linewidth]{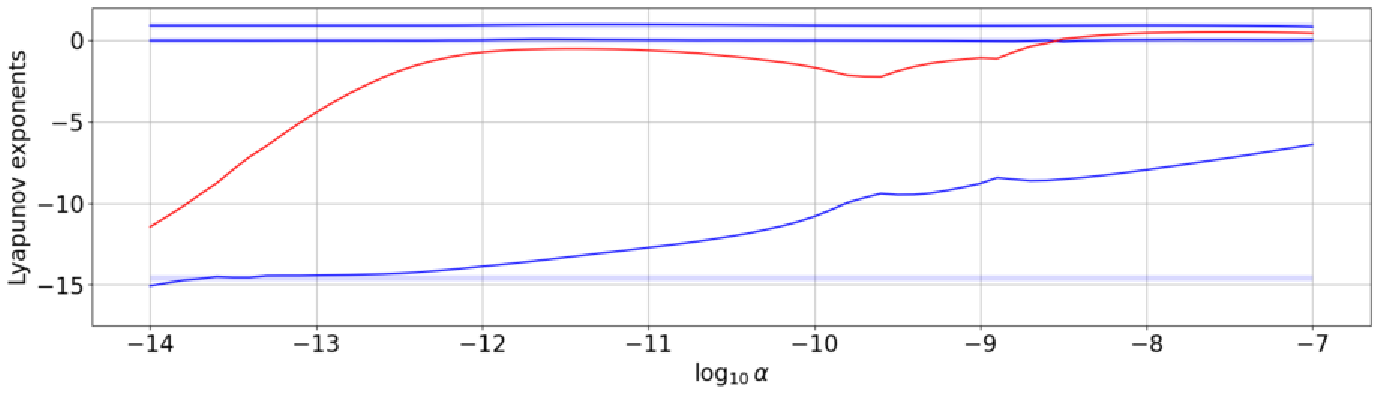}
    \includegraphics[width=0.45\linewidth]{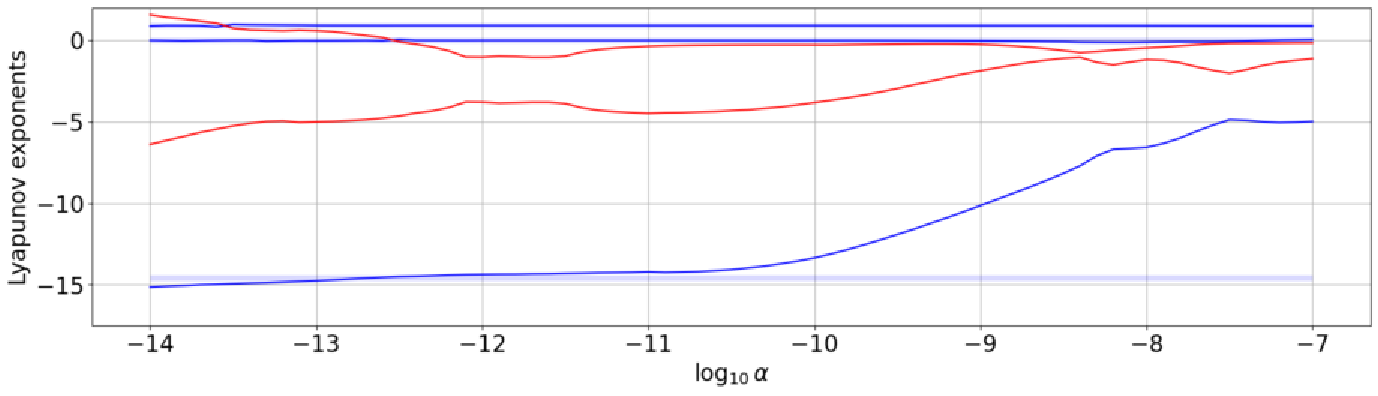}
    \caption{
    {\bf Lyapunov spectra of RfR models for the Lorenz system with different embedding dimensions.}
    {\it Top:} $D=4$ with $\tau=0.07$.
    {\it Bottom:} $D=5$ with $\tau=0.04$.
    The delay time $\tau$ is adjusted depending on the embedding dimension so that the amount of information contained in the delay coordinates is comparable.
    As in the H\'enon case, the physically dominant Lyapunov exponents are robust, whereas the spurious exponents are sensitive to the hyperparameter, and the same qualitative behavior is observed across different embedding dimensions.
}
\label{fig:add-results-lorenz}
\end{figure}

To examine the dependence of the results on the embedding dimension, we performed additional experiments using different embedding dimensions for both the H\'enon map and the Lorenz system.
Figure~\ref{fig:add-results-henon} shows the results for the H\'enon map with embedding dimensions $D=3$ (left) and $D=5$ (right).
Figure~\ref{fig:add-results-lorenz} shows the results for the Lorenz system with $D=4$ (left) and $D=5$ (right). 
For the Lorenz system, the delay time $\tau$ is adjusted depending on the embedding dimension so that the amount of information contained in the delay coordinates is comparable. 
In particular, smaller embedding dimensions require larger delay times.

In all cases, we observe that the physically dominant Lyapunov exponents remain robust with respect to the hyperparameter in a certain range, whereas the spurious exponents vary sensitively.
This qualitative structure is consistent across different embedding dimensions, supporting that the conclusions of the main text do not depend on a specific choice of embedding dimension.

\end{document}